\documentclass[pra,aps,twocolumn,superscriptaddress,notitlepage]{revtex4-1}
\usepackage{amssymb,amsfonts,amsmath,amstext,graphicx,hyperref}
\usepackage{tikz}
\usepackage{graphicx}
\usepackage[normalem]{ulem}
\hypersetup{
	colorlinks=true,
	linkcolor=blue,
	filecolor=magenta,      
	urlcolor=cyan,
}

\def\bra#1{\langle{#1}|}
\def\ket#1{|{#1}\rangle}

\begin{document}

\title{Generalized Measure of Quantum synchronization}
\author{Noufal Jaseem}
\email{noufal@iitb.ac.in}
\affiliation{Department of Physics, Indian Institute of Technology-Bombay, Powai, Mumbai 400076, India}

\author{Michal Hajdu\v{s}ek}
\email{michal@sfc.wide.ad.jp}
\affiliation{Keio University Shonan Fujisawa Campus, 5322 Endo, Fujisawa, Kanagawa 252-0882, Japan}

\author{Parvinder Solanki}
\affiliation{Department of Physics, Indian Institute of Technology-Bombay, Powai, Mumbai 400076, India}

\author{Leong-Chuan Kwek}
\affiliation{Centre for Quantum Technologies, National University of Singapore, 3 Science Drive 2, 117543 Singapore, Singapore}
\affiliation{Institute of Advanced Studies, Nanyang Technological University, Singapore 639673}
\affiliation{National Institute of Education, Nanyang Technological University, Singapore 637616}

\author{Rosario Fazio}
\affiliation{ICTP, Strada Costiera 11, 34151 Trieste, Italy}
\affiliation{Dipartimento di Fisica , Università di Napoli Federico II, Complesso di Monte S. Angelo, 80126 - Napoli, Italy}
\thanks{On leave}

\author{Sai Vinjanampathy}
\email{saiv@phy.iitb.ac.in}
\affiliation{Department of Physics, Indian Institute of Technology-Bombay, Powai, Mumbai 400076, India}
\affiliation{Centre for Quantum Technologies, National University of Singapore, 3 Science Drive 2, 117543 Singapore, Singapore}

\date{\today}

\begin{abstract}
We present a generalized information-theoretic measure of synchronization in quantum systems.
This measure is applicable to dynamics of anharmonic oscillators, few-level atoms, and coupled oscillator networks. Furthermore, the new measure allows us to discuss synchronization of disparate physical systems such as coupled hybrid quantum systems and coupled systems undergoing mutual synchronization that are also driven locally. In many cases of interest, we find a closed-form expression for the proposed measure.
\end{abstract} 
\maketitle

\section{Introduction \label{sec:introduction}}

Detecting and measuring synchronization of classical systems has been an important area of research in nonlinear dynamics for decades \cite{strogatz2004sync,pikovsky2003synchronization,balanov2008synchronization}.
Regardless of the exact nature of the studied classical system, be it a collection of fireflies, heart cells or firing neurons in the brain, the equations of motion generate trajectories in phase space which enable us to compute an appropriate measure of synchronization.

Quantification of synchronization in quantum systems, on the other hand, does not immediately present such a unified and intuitive approach.
Different measures have been introduced to study externally driven and mutually coupled van der Pol oscillators \cite{walter2014quantum,lee2013quantum,sonar2018squeezing,lee2014entanglement,walter2015quantum,bastidas2015quantum}, driven and coupled spin-1 atoms \cite{roulet2018synchronizing,roulet2018quantum}, interacting many-body systems \cite{witthaut2017classical}, spins interacting via coupled optical cavities \cite{ameri2015mutual}, coupled opto-mechanical systems \cite{mari2013measures,weiss2016noise}, and quantum systems undergoing transient synchronization \cite{manzano2013synchronization,giorgi2019transient}. Furthermore, recently such quantum synchronization has been experimentally observed \cite{laskar2019observation,PhysRevResearch.2.023026}.
\textcolor{black}{These approaches are in most cases tailored on specific systems.
Therefore it would be very interesting to seek a generic quantifier.}

In this manuscript, we address this issue by proposing a new measure of quantum synchronization that builds upon insight from quantum information while bringing in several desirable new attributes.
We exploit the approach that in order to quantify certain quality of a state (eg. entanglement) one can compute the distance to the nearest state lacking this quality (eg. set of separable states).
This strategy has been extremely fruitful in several contexts such as quantification of entanglement \cite{vedral1997quantifying,hajdusek2010entanglement,hajdusek2013direct}, discord \cite{modi2010unified,modi2012classical}, and quantum coherence \cite{baumgratz2014quantifying}.
We introduce the \textit{set of unsynchronized limit-cycle states}, the distance to which is then optimized in order to quantify synchronization.
\textcolor{black}{They will fulfil a similar role that separable states play in entanglement quantification or that incoherent states play in coherence quantifiction.
Note that the unsynchornized limit-cycle states are conceptually different from the notion of a semi-classical limit cycle state used in \cite{lee2013quantum,walter2014quantum}.}
We show that this approach can be immediately applied in the case of both a single quantum system entrained to an external signal as well as a number of systems undergoing mutual synchronization.
Furthermore, we prove that our synchronization measure works for finite and infinite-dimensional systems.
Finally, we demonstrate that by construction the limit-cycle state allows for great flexibility and can even be applied to study synchronization of disparate quantum systems.
\begin{figure}[h]
	\includegraphics[width=0.95\linewidth]{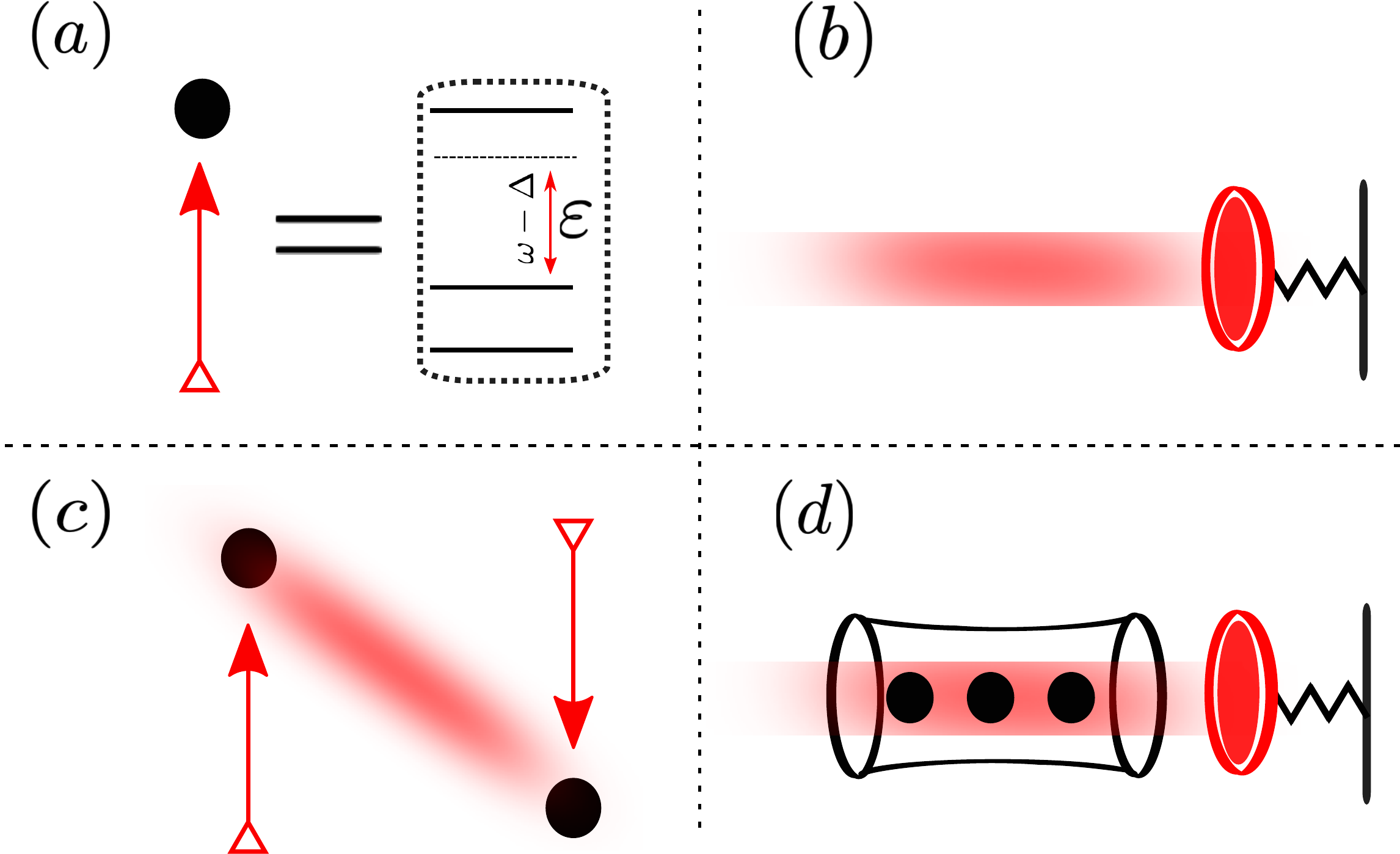}
	\caption{\label{fig:illustration} Illustrative diagrams: A glimpse of the variety of systems that can be studied using the distance-based measure introduced in this paper. a) Driven 3-level system. b) Driven quantum vdP. c) Two driven 3-level systems which are coupled to each other. d) quantum vdP coupled to mutually coupled 3-level systems.}
\end{figure}

\textcolor{black}{In the next section we define our new measure of synchronization.
In Section \ref{sec:examples} we review two canonical examples of systems studied in quantum synchronization.
In Section \ref{sec:unipartite} and Section \ref{sec:bipartite}, we apply the unified measure to study unipartite and bipartite systems, respectively with relevant examples.
Section \ref{sec:discussion} provides a concluding discussion.}

\section{Relative entropy of synchronization \label{sec:new_measure}}

\textcolor{black}{In analogy to classical dynamical systems, synchronization in quantum systems begins by establishing the existence of a limit-cycle.
In classical mechanics, establishing a limit-cycle implies that the relevant free coordinates are identified, which in the case of a van der Pol oscillator for example is its phase \cite{pikovsky2003synchronization}.
Once the limit-cycle is established, the dynamical systems may be (a) entrained to an external frequency standard or (b) coupled with each other in order to observe mutual synchronization.}

The common physical insight that goes into defining the different measures of synchronization can be summarised as (a) identifying the limit-cycle behavior in the quantum system and (b) quantifying deviations from such a state.
Motivated by the current discussion, a natural choice of quantifying this deviation is by considering the distance $\mathfrak{D}(\rho, \rho_{\text{lim}})$ between the steady state $\rho$ of the evolution and its limit-cycle state $\rho_{\text{lim}}$.
The distance $\mathfrak{D}$ in fact does not need to be a strict measure but can also be distance-like as we will demonstrate in this section.

\begin{figure}[t]
	\includegraphics[width=\linewidth]{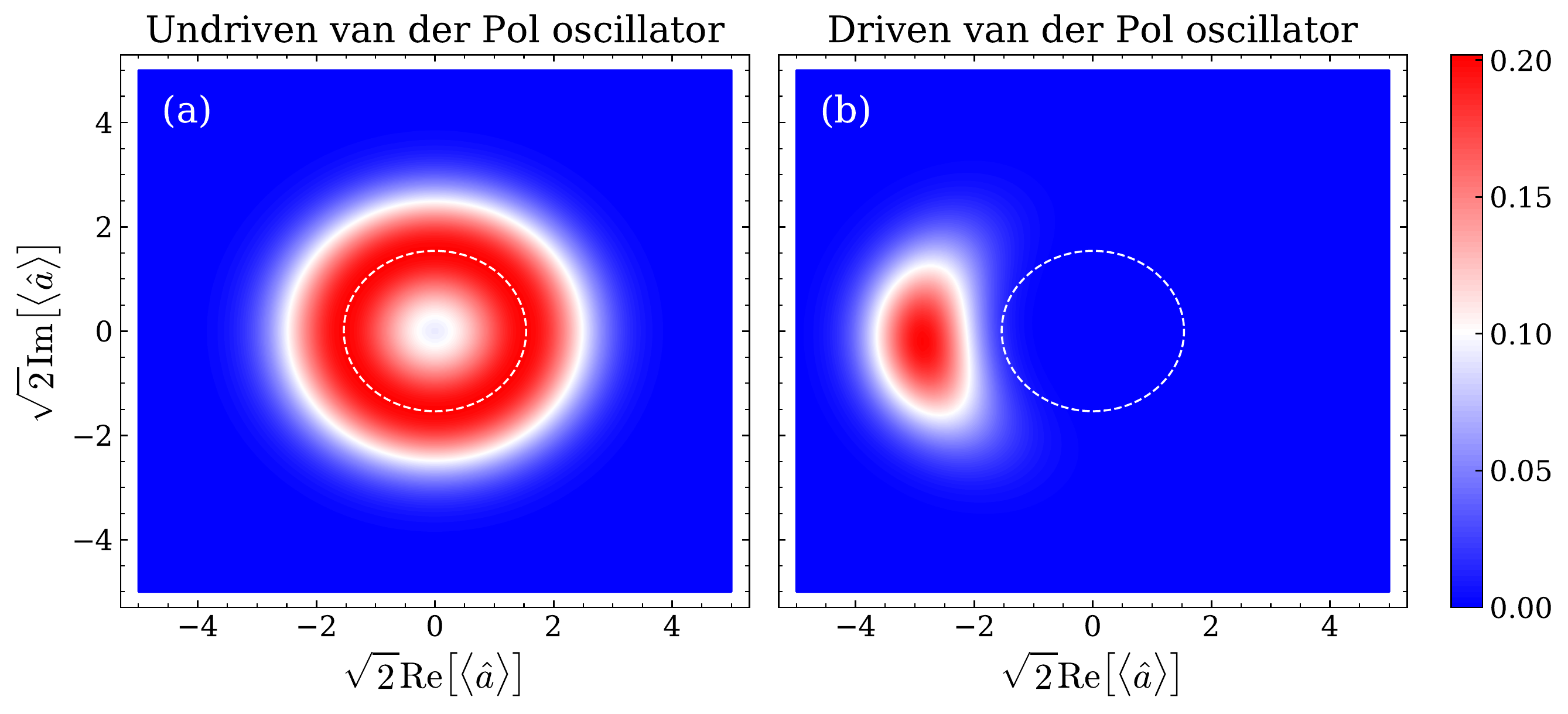}
	\caption{\label{fig:figure2} (a) Wigner function of the steady state of an undriven van der Pol oscillator of Eq.~(\ref{eq:vdp_driven}). Parameters used are $\Delta/\Gamma_g=0.1$, $\Gamma_d/\Gamma_g=0.5$ and $\varepsilon/\Gamma_g=0$. (b) The external drive entrains the oscillator, as seen by the localization of the Wigner function, as well as changes its population distribution. The white dashed circle in (a) and (b) depicts the average population of an undriven oscillator. Parameters are the same except for $\varepsilon/\Gamma_g=3$. Note that the driving strength has been set this high in order to demonstrate the effect of population change clearly. Synchronization is usually limited to perturbative driving where the effect of population change is much smaller.}
\end{figure}

A question that remains is what is the appropriate limit-cycle state $\rho_{\text{lim}}$ that can be used to quantify synchronization.
In general, it is not the steady state of the evolution $\dot{\rho}=\mathcal{L}[\rho]$ where the external drive $\varepsilon$ and the mutual coupling $g$ have been switched off.
This is because the external drive and mutual coupling have two disparate effects.
First, they produce entrainment/synchronization that is of interest to us.
Besides this, they also may change the population of the system which does not affect the synchronization properties of the system.
Proper measure of synchronization should quantify only the first effect and be insensitive to the latter. We demonstrate these two effects schematically on an externally driven van der Pol oscillator in Fig.~\ref{fig:figure2}.

In order to quantify the synchronization of a steady state $\rho$ we must minimize the distance $\mathfrak{D}$ over all possible limit-cycle states, leading to the definition of our new measure of synchronization,
\begin{equation}
    \Omega(\rho) \equiv \min_{\sigma\in \Sigma} \mathfrak{D}(\rho, \sigma),
    \label{eq:measure_new}
\end{equation}
where $\Sigma$ is the set of all possible limit-cycle states.

We consider two important examples of distance functions $\mathfrak{D}$.
The first one is the relative entropy of synchronization,
\begin{equation}
    \Omega_{\text{R}}(\rho) \equiv \min_{\sigma\in\Sigma} S(\rho || \sigma),
    \label{eq:measure_relent}
\end{equation}
where $S(\rho || \sigma)=\text{Tr}[\rho\log\rho - \rho\log\sigma]$.
This measure is suitable when $\sigma$ is full rank which is true in the cases that we consider.
In order to handle limit-cycles which are not full rank we propose the trace distance to quantify synchronization,
\begin{equation}
    \Omega_{\text{D}}(\rho) \equiv \min_{\sigma\in\Sigma} \|\rho - \sigma\|_1,
    \label{eq:measure_trace}
\end{equation}
where $\|O\|_1=\text{Tr}[\sqrt{O^{\dagger}O}]$.

The set of limit-cycle states $\Sigma$ depends on two things.
First, it depends on the particular dynamics under investigation.
The form of the limit-cycle state $\rho_{\text{lim}}$ will be different for a single system driven externally and for two coupled systems undergoing mutual synchronization.
Second, we might be interested in more complex scenarios where for instance two coupled systems are also being locally driven.
In such cases the limit-cycle states require a suitable form that reflects not only the dynamics being studied but also that distinguishes between synchronization due to mutual coupling and entrainment due to the local drives.

In order for our proposed measure in Eq.~(\ref{eq:measure_new}) to be a physically reasonable quantity the form of the limit-cycle states must be such that the already existing measures of synchronization vanish for these states.
This will be the guiding principle in defining three classes of limit-cycle states which we study.
Two of these classes are inspired by existing results while the third class is newly defined to specifically demonstrate the flexibility of our proposed measure.

\textit{Diagonal limit-cycle states.---}
It follows from the definition of a free observable phase that a self-sustaining system which is not driven and not coupled to other systems will be diagonal in the energy eigenbasis.
Diagonal limit-cycle states are useful when studying entrainment of a single quantum system to an external drive \cite{walter2014quantum,roulet2018synchronizing} as well as mutual synchronization of coupled systems \cite{weiss2016noise,roulet2018quantum}.
They are physically motivated by the fact that existing measures all vanish for diagonal states.
Furthermore their simple form allows us to find a closed form for the minimization in Eq.~(\ref{eq:measure_new}).

\textit{Marginal limit-cycle states.---}
Mutual information is well defined both in the classical and quantum world, making it an appropriate measure of quantum systems which can exhibit syncrhonization both in the classical and quantum regime.
Motivated by this, authors in \cite{ameri2015mutual} introduced this class of limit-cycle states to study coupled van der Pol oscillators.
Marginal limit-cycle states do not necessarily need to be diagonal making them more general when compared with diagonal limit-cycle states.
They were motivated physically showing that the mutual information measure vanishes when measure of complete synchronization in \cite{ameri2015mutual} does as well.

\textit{Partially-coherent limit-cycle states.---} This previously unstudied class of limit-cycle states is the third example considered in this manuscript. These states offer a large degree of flexibility to the form of the limit-cycle state over which one minimizes the distance measures and is particularly suited to situations where there are a number of sources that may entrain or synchronize a system and our aim is to isolate a subset of these sources.
Therefore they are even more general than marginal limit-cycle states since they offer direct control over the coherences present in the state. An example of such scenario is given in Eq.~(\ref{eq:coupled_driven_spins}) by two coupled spin-1 atoms which are also locally driven.

In the following sections we demonstrate how these three classes of limit-cycle states are used in various scenarios and derive some useful properties of the new measure of synchronization.

\section{Examples of synchronizing systems \label{sec:examples}}


The quantum systems undergoing synchronization can be split into two groups, continuous variable systems and finite-dimensional systems.
We focus on one canonical representatives from each group.
In this section we outline the van der Pol oscillator, a continuous variable model, while the finite-dimensional systems are represented by a spin-1 atom.
Both of these systems can be used to study entrainment of a single oscillator to an external drive as well as mutual synchronization of a number of coupled oscillators.

\textit{Continuous variable systems.---} Infinite-dimensional systems are a natural candidate to study synchronization since they follow in close analogy from their classical counterparts. Following this intuition, the authors in \cite{walter2014quantum,lee2013quantum} studied a driven harmonic oscillator mode undergoing Lindblad dynamics due to two baths,
\begin{equation}
	\dot{\rho} = -i[\hat{H},\rho] + \Gamma_g \mathcal{D}[\hat{a}^\dagger]\rho + \Gamma_d \mathcal{D}[\hat{a}^2]\rho.
	\label{eq:vdp_driven}
\end{equation}
Here $\hat{H} = -\Delta \hat{a}^\dagger \hat{a} + i\varepsilon(\hat{a} - \hat{a}^\dagger)$ with $\Delta$ being the detuning of the drive's frequency $\omega_{d}$ from the natural frequency of the oscillator $\omega_0$, and $\mathcal{D}[\hat{O}]\rho =( 2 \hat{O} \rho \hat{O}^{\dagger}- \hat{O}^\dagger \hat{O} \rho - \rho \hat{O}^\dagger \hat{O})/2$.
Note that just like classical dynamical systems require a source and sink of energy to compete in a nonlinear manner in order to form a limit-cycle, Eq.~(\ref{eq:vdp_driven}) has two dissipators, one that removes two excitations incoherently from the system and another that adds one excitation incoherently to the system.
This ensures that the amplitude of oscillations is stable and the quantum van der Pol oscillator possesses a limit cycle.

Two van der Pol oscillators can be coupled in a number of possible ways.
Authors in \cite{lee2013quantum} considered a coherent coupling by adding an interaction Hamiltonian,
\begin{equation}
    \hat{H}_{\text{int}} = g \left( \hat{a}_1^{\dagger}\hat{a}_2 + \hat{a}_1\hat{a}_2^{\dagger} \right).
    \label{eq:coupling_vdP_coherent}
\end{equation}
Another possibility is to couple the two oscillators dissipatively by adding another Lindblad operator as in \cite{walter2015quantum,eshaqi-sani2020synchronizing},
\begin{equation}
    \dot{\rho} = \mathcal{L}_1[\rho] + \mathcal{L}_2[\rho] + g\mathcal{D}[\hat{a}_1-\hat{a}_2]\rho,
    \label{eq:coupling_vdP_dissipative}
\end{equation}
where $\mathcal{L}_i[\rho]$ is the Lindblad evolution given by Eq.~(\ref{eq:vdp_driven}) with $\varepsilon=0$.

\textit{Finite-level systems.---} 
Investigation of synchronisation in the quantum regime motivated the study of small quantum systems with no classical analogs.
Authors in \cite{roulet2018synchronizing} explored an externally driven spin-1 atom.
In the frame rotating with the external signal,
\begin{equation}
	\dot{\rho} = -i[\hat{H}, \rho] + \frac{\gamma_g}{2} \mathcal{D}[\hat{S}_+ \hat{S}_z]\rho + \frac{\gamma_d}{2} \mathcal{D}[\hat{S}_- \hat{S}_z]\rho,
	\label{eq:three_level}
\end{equation}
where $\hat{H} = \Delta\hat{S}_z + \varepsilon \hat{S}_y$ and $\hat{S}_{\pm}$ are the raising/lowering operators.
The limit cycle is established by the nonlinear nature of the dissipator operators stabilizing the middle energy level.
This concentrates the whole atomic population into the middle level meaning the limit cycle state is pure unlike in the case of a van der Pol oscillator.
General framework to study entrainment of spin-1 atoms to an external signal was presented in \cite{koppenhofer2019optimal}.


Mutual synchronization of two such spin-1 systems was considered in \cite{roulet2018quantum} by introducing a coherent interaction term,
\begin{equation}
    \hat{H}_{\text{int}} = i g \left( \hat{S}_-^A\hat{S}_+^B - \hat{S}_+^A\hat{S}_-^B \right),
    \label{eq:coupling_spin_coherent}
\end{equation}
in the context of entanglement production in the steady state.

In order to demonstrate the flexibility of our proposed new measure of synchronization we also consider a system of two coherently coupled spin-1 atoms where each subsystem is also driven locally,
\begin{equation}
    \dot{\rho} = -i[\hat{H}, \rho] + \sum_{\alpha=A,B} \left\{\frac{\gamma_g^{\alpha}}{2}\mathcal{D}[\hat{S}_+^{\alpha}\hat{S}_z^{\alpha}] + \frac{\gamma_d^{\alpha}}{2}\mathcal{D}[\hat{S}_-^{\alpha}\hat{S}_z^{\alpha}] \right\} \rho,
    \label{eq:coupled_driven_spins}
\end{equation}
where the coherent part of the evolution is given by
\begin{align}
    \hat{H} = \delta\hat{S}_z^A + (\delta+\Delta)\hat{S}_z^B & + \epsilon\sum_{\alpha=A,B} \left( \hat{S}_z^{\alpha}\hat{S}_+^{\alpha} + \hat{S}_-^{\alpha}\hat{S}_z^{\alpha} \right) \nonumber\\
    & + ig\left( \hat{S}_+^A\hat{S}_-^B - \hat{S}_-^A\hat{S}_+^B \right), 
\end{align}
with $\delta$ being the detuning of the local drives from the natural frequencies of the spin-1 atoms and $\Delta$ the detuning between the atoms.

\section{Unipartite Systems \label{sec:unipartite}}

Let us consider the scenario where a single quantum system is entrained to an external drive \cite{walter2014quantum, lee2013quantum}.
In the absence of the external drive the steady state of the evolution is diagonal in the energy eigenbasis $\{|E_i\rangle\}$.
Introducing the external drive has two consequences.
First, it produces a change in the populations given by the diagonal elements of the steady-state density matrix.
Second, it populates the off-diagonal elements of the steady-state density matrix $\rho$ and may produce entrainment to the external drive.

Consider a diagonal limit-cycle state,
\begin{equation}
    \rho_{\text{lim}} = \sum_j q_j |E_j\rangle\langle E_j|.
    \label{eq:limit_diagonal}
\end{equation}
The relative entropy $S(\rho || \rho_{lim})$ can be expanded as
\begin{equation}
	S(\rho || \rho_{\text{lim}}) = S_{\text{coh}}(\rho) + D_{KL}[p||q],
	\label{eq:relative_entropy}
\end{equation}
where $S_{\text{coh}}(\rho)=S(\rho_{\text{diag}})-S(\rho)$ is the relative entropy of coherence \cite{baumgratz2014quantifying}, $S(\rho)=-\text{Tr}[\rho\log\rho]$ is the von Neumann entropy, and $\rho_{\text{diag}}$ is obtained from $\rho$ by setting its off-diagonal terms to zero.
The second term in Eq.~(\ref{eq:relative_entropy}) given by the Kullback-Leibler divergence of populations of the steady state $\rho$ and the limit-cycle state $\rho_{\text{lim}}$ describes the change in the population of the driven system mentioned in Section \ref{sec:new_measure} and pictured in Fig.~\ref{fig:figure2}.

\begin{figure}[t]
	\includegraphics[width=0.95\linewidth]{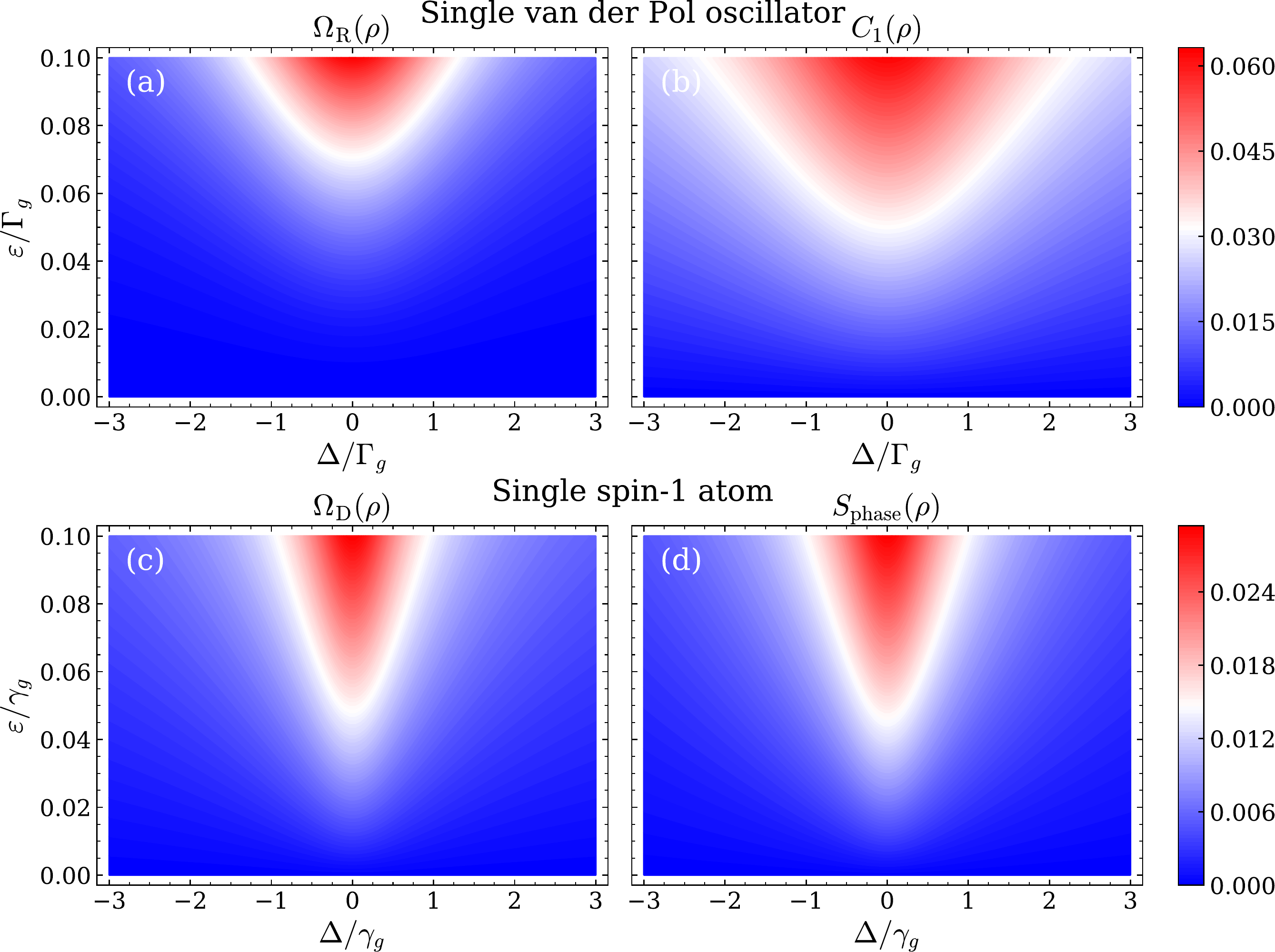}
	\caption{\label{fig:figure3} (a) Relative entropy of synchronization $\Omega_{\text{R}}(\rho)$ of an externally driven van der Pol oscillator displays qualitatively same behavior as $C_1(\rho)$ measure of synchronization of \cite{weiss2016noise} shown in (b). The damping rates are $\Gamma_d/\Gamma_g=10$. (c) Trace distance measure of synchronization $\Omega_{\text{D}}(\rho)$ for an externally driven spin-1 atom. (d) Phase-locking measure $S_{\text{phase}}(\rho)$ of \cite{roulet2018synchronizing}. The damping rates for the spin-1 atom are $\gamma_d/\gamma_g=10$.}
\end{figure}

The minimization of Eq.~(\ref{eq:relative_entropy}) over all diagonal limit-cycle states can be done simply by setting $\rho_{\text{lim}}=\rho_{\text{diag}}$,
\begin{equation}
    \Omega_{\text{R}}(\rho) = S_{\text{coh}}(\rho).
    \label{eq:sync_coh_relent}
\end{equation}
Eq.~(\ref{eq:sync_coh_relent}) provides a closed-form expression for the minimization problem in Eq.~(\ref{eq:measure_relent}) and shows that relative entropy of synchronization and relative entropy of coherence are equivalent when considering a single quantum system driven externally.
In Fig.~\ref{fig:figure3}(a) we plot $\Omega_{\text{R}}(\rho)$ of a harmonically driven van der Pol oscillator by numerically solving Eq.~(\ref{eq:vdp_driven}) for its steady state using QuTip \cite{johansson2012qutip1,johansson2013qutip2}.
We observe the usual Arnold tongue as expected from a well-behaved measure of synchronization.
\textcolor{black}{In Fig.~\ref{fig:figure3}(b), we compare this against $C_1(\rho)=|\langle \hat{a} \rangle| / \sqrt{\langle\hat{a}^{\dagger}\hat{a}\rangle}$, which is a synchronization measure \cite{weiss2016noise}.}

Using a similar strategy to perform the minimization we can find an upper bound on the trace distance measure of synchronization $\Omega_{\text{D}}$.
The trace distance satisfies the following inequality,
\begin{equation}
    \|\rho - \rho_{\text{lim}}\|_1 \leq \sum_j |p_j - q_j| + \sum_{j\neq k}|\rho_{jk}|.
    \label{eq:trace_dist_ineq}
\end{equation}
Setting $\rho_{\text{lim}}=\rho_{\text{diag}}$ in Eq.~(\ref{eq:trace_dist_ineq}) we obtain an upper bound on the trace distance measure of synchronization,
\begin{equation}
    \Omega_{\text{D}}(\rho) \leq C_{l_{1}}(\rho),
    \label{eq:sync_trace_dist_ineq}
\end{equation}
where $C_{l_{1}}(\rho)=\sum_{j\neq k}|\rho_{jk}|$ is the $l_1$-norm of coherence introduced in \cite{baumgratz2014quantifying}.
Fig.~\ref{fig:figure3}(c) shows the Arnold tongue of $\Omega_{\text{D}}(\rho)$ for a driven spin-1 atom model of Eq.~(\ref{eq:three_level}).
\textcolor{black}{For reference we display the phase measure of synchronization $S_{\text{phase}}(\rho)$ defined in Eq.~(8) of \cite{roulet2018synchronizing}.}

The connection between synchronization and coherence has been noted previously in literature, for example in \cite{koppenhofer2019optimal, jaseem2020quantum}.
In fact, for some existing measures one can quickly see that they vanish when the steady state $\rho$ is diagonal in the energy eigenbasis.
The result of Eq.~(\ref{eq:sync_coh_relent}) is the first time when it was shown that a measure of synchronization reduces exactly to one of the well-established measures of coherence.
On the other hand we will show in the next section that this is not the case anymore for general bipartite synchronization, demonstrating that synchronization is generally not equivalent to coherence. 

\section{Bipartite systems \label{sec:bipartite}}

In this section, we apply our measure to bipartite systems and study their mutual synchronization.
The minimization in Eq.~(\ref{eq:measure_relent}) for general limit-cycle states $\sigma=\sum_i q_i\sigma_i^A\otimes\sigma_i^B$, where $\sigma^{A/B}$ are the limit-cycle states for the individual subsystems, proceeds as follows,
\begin{eqnarray}
    \Omega_{\text{R}}(\rho) & = & \min_{\sigma \in \Sigma} S(\rho||\sigma) \nonumber\\
    & = & -S(\rho) - \max_{q_i,\sigma_i^A,\sigma_i^B} \;\text{Tr}[\rho \log (\sum_i q_i \sigma_i^A\otimes\sigma_i^B)]\nonumber\\
    & \leq & -S(\rho) - \max_{q_i,\sigma_{i}^A} \;\sum_i q_{i} \text{Tr}[\rho_A \log \sigma_i^{A}] \nonumber\\
    & & \qquad - \max_{q_i,\sigma_i^{B}} \;\sum_i q_{i} \text{Tr} [\rho_B \log \sigma_i^{B}]\nonumber\\
    & \leq & -S(\rho) -\omega_A - \omega_B,
    \label{eq:inequality}
\end{eqnarray}
where we defined $\omega_{\alpha} = \max_{q_i,\sigma^{\alpha}_i} \;\sum_i q_{i} \text{Tr} [\rho_{\alpha} \log \sigma^{\alpha}_i]$.
In the above, we have used the concavity of the expression, $\text{Tr}[\rho \log (\sum_i q_i \sigma_i)]$ as $\text{Tr}[\rho \log (\sum_i q_i \sigma_i)] \geq \sum_i q_i \text{Tr} [\rho \log \sigma_i]$ for $0<q_i<1$ and $\sum_i q_i =1$. This inequality saturates when $q_1=1$ and the rest of $q_i$'s are zero.
Though analytically evaluating $\omega_{\alpha}$ is difficult for general limit-cycle states, we show in the rest of this section that for many classes of limit-cycle states it is possible to find closed form expressions for the minimization in Eq.~(\ref{eq:measure_relent}).

\begin{figure*}[t]
    \centering
    \begin{tikzpicture}
        \node[anchor=south west,inner sep=0] (image) at (0,0) {\includegraphics[width=\textwidth]{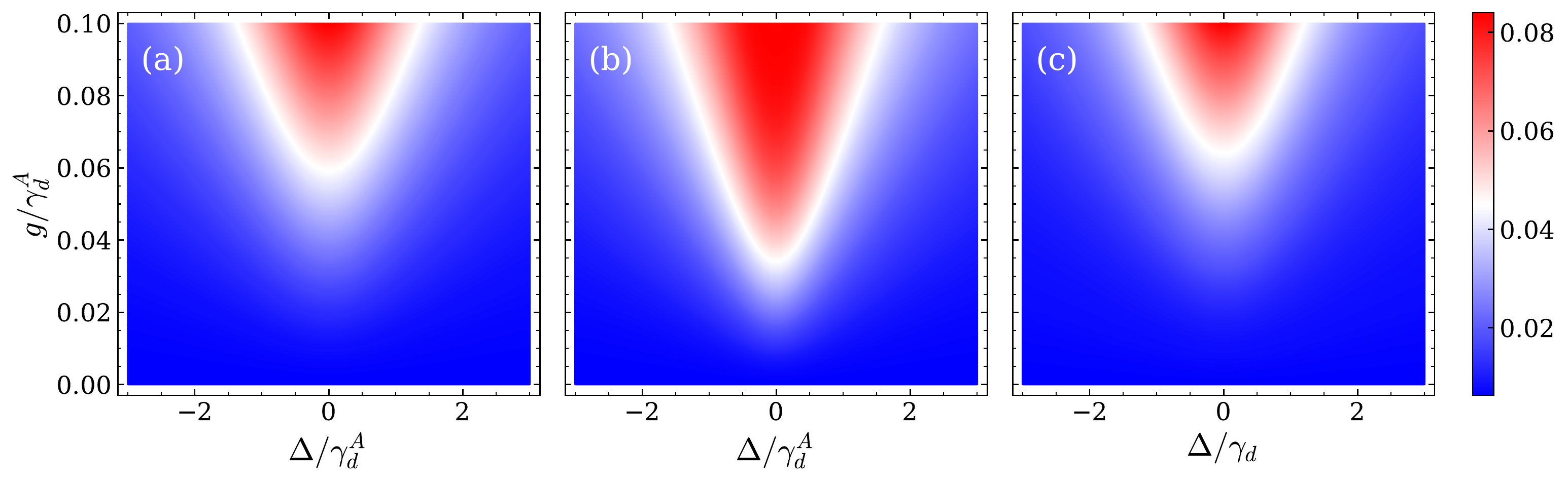}};
        \begin{scope}[x={(image.south east)},y={(image.north west)}]
            \node[anchor=south west,inner sep=0] (image) at (0.65,0.2) {\includegraphics[width=0.135\textwidth,height=1.6cm]{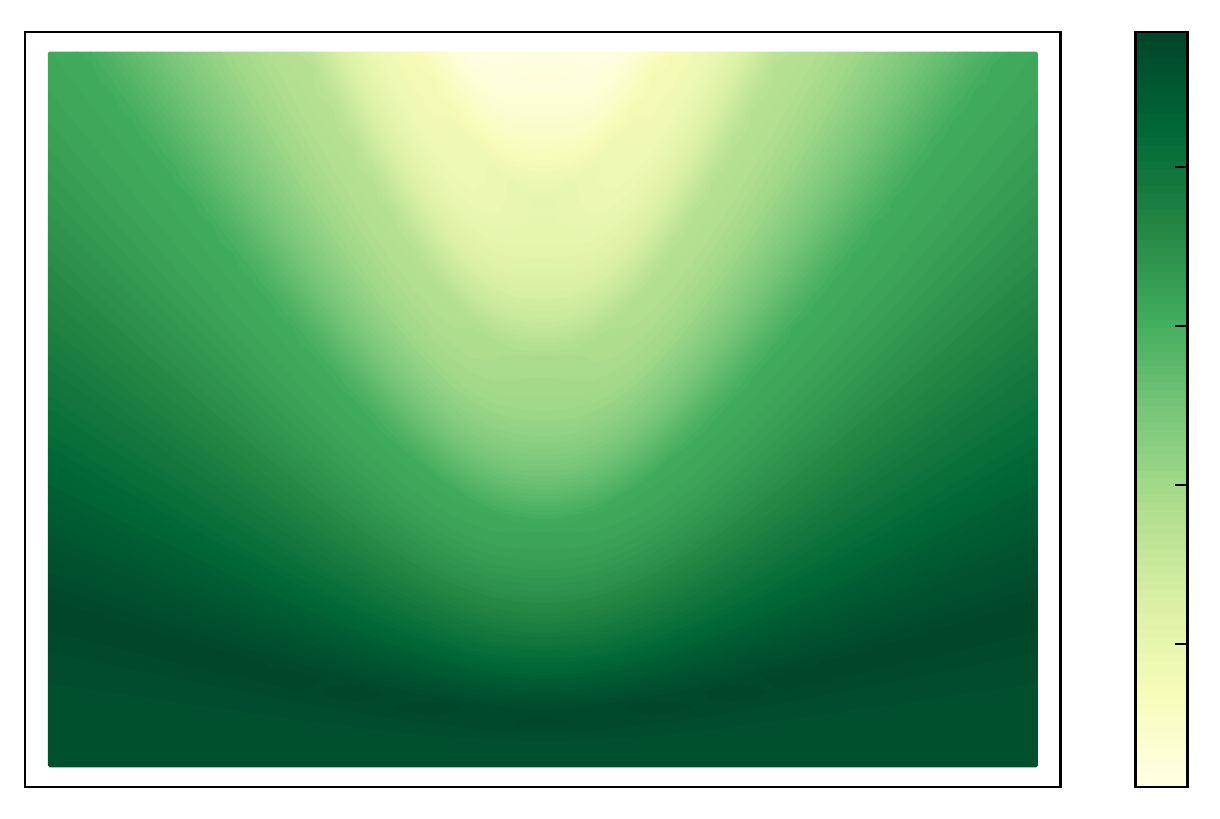}};
        \end{scope}
    \end{tikzpicture}
    \caption{\label{fig:figure4} Relative entropy of synchronization $\Omega_{\text{R}}(\rho)$ for two mutually coupled systems. (a) Two spin-1 atoms coupled coherently as in Eq.~(\ref{eq:coupling_spin_coherent}). The dissipators are $\gamma_g^A/\gamma_d^A=\gamma_d^B/\gamma_d^A=100$ and $\gamma_g^B=\gamma_d^A$. (b) Spin-1 atom coupled to a van der Pol oscillator as in Eq.~(\ref{eq:vdP_spin}). The parameters are $\Gamma_d/\gamma_d=0.1$, $\gamma_g/\gamma_d=100$, and $\Gamma_g=\gamma_d$ (c). Two coupled spin-1 atoms that are also driven locally as in Eq.~(\ref{eq:coupled_driven_spins}). The dissipators are the same as in subfigure (a), the local drives are resonant with the atoms, $\delta/\gamma_d^A=0$, and their strength is $\varepsilon/\gamma_d^A=0.01$. The effect of the local drives is to decrease mutual synchronization between the atoms. Qualitative difference between $\Omega_R$ for partially-coherent limit-cycles and the mutual information is shown in the inset.}
\end{figure*}

\subsection{Diagonal limit-cycle states \label{sec:diagonal}}

Owing to classical correlations, the set of diagonal limit cycle states denoted as $\delta$ is more general than just the tensor product of diagonal states. The relative entropy of synchronization for the set of diagonal limit-cycle states can be evaluated to be
\begin{eqnarray}
    \label{eq:coh_meas}
    \Omega_{\text{R}}(\rho) & = & \min_{\delta} S(\rho||\delta) \nonumber\\
    & = & \min_{\delta} \left[ S_{\text{coh}}(\rho) + S(\rho_{\text{diag}}||\delta) \right] \nonumber\\
    & = & S_{\text{coh}}(\rho).
\end{eqnarray}
The minimization of the relative entropy of synchronization over $\delta$ yields the relative entropy of coherence, similar to our discussion in the unipartite case.
This result is obtained when the set of diagonal limit-cycle states is not restricted to a simple tensor product.
Convex mixtures of diagonal limit-cycle states are themselves diagonal-limit cycle states and hence they do not produce any synchronization.
The physical motivation for including these convex mixtures of limit-cycle states is that if two quantum systems are coupled via operators that are diagonal with respect to the eigenbasis of the undriven steady states (either in the Hamiltonian or Lindbladians), while these couplings can give rise to correlations, they cannot generate synchronization.
We account for the buildup of such classical correlations in this scenario by optimizing over the full set of diagonal states.

To further illustrate this point, consider a scenario where the limit-cycle states are composed of uncorrelated diagonal states, namely $\delta=\delta^A\otimes\delta^B$.
Such a physical scenario arises when two hitherto non-interacting quantum systems are coupled by a \textit{synchronizing} coupling either in the Hamiltonian \cite{lee2013quantum} or Lindbladians \cite{walter2015quantum}.
Since they are non-interacting without the synchronizing coupling, their limit-cycle states will be uncorrelated.
Minimization over such local diagonal limit-cycle states yields,
\begin{equation}
    \min_{\delta^A,\delta^B} S(\rho || \delta^A\otimes\delta^B) = S_{\text{coh}}(\rho) + I_c(\rho),
\end{equation}
where $I_c(\rho)=S(\rho_{\text{diag}}^A)+S(\rho_{\text{diag}}^B)-S(\rho_{\text{diag}})$ is the classical mutual information.
Since $I_c(\rho)\geq0$, we see that the additional classical correlations developed by the uncoupled systems are properly accounted for in the synchronization measure.
This same term $I_c(\rho)$ is removed when the limit-cycles are classically correlated in Eq.~(\ref{eq:coh_meas})

As an example, we considered two mutually coupled 3-levels systems that were studied in \cite{roulet2018quantum}.
The corresponding relative entropy of synchronization is depicted in the Fig.~\ref{fig:figure4}(a), which shows agreement with known results.
 
\subsection{Synchronization of dissimilar systems \label{sec:dissimilar}}

Until now, the study of mutual synchronization has been restricted to coupling identical systems~\cite{ameri2015mutual,walter2015quantum,lee2013quantum}. In contrast to this, quantum technology platforms such as quantum optics, cavity and circuit quantum electrodynamics and many-body physics routinely deal with coupling dissimilar systems. As a prototypical example, let us consider the Tavis-Cummings model whose Hamiltonian within rotating wave approximation is given by 
\begin{equation}
    \hat{H}=\Omega \hat{J}_z +\omega (\hat{a}^{\dagger}\hat{a}+\frac{1}{2})+\epsilon(\hat{a}\hat{J}_{+}+\hat{a}^{\dag}\hat{J}_{-}).    
\end{equation}
Here $\hat{J}_i=\sum_\alpha \sigma^{(i)}_{\alpha}/2$ are the collective spins of $N$ atoms and $\hat{a}$ is a bosonic annihilation operator.
In the large $N$ limit, this Hamiltonian could be transformed into a coupled bosonic Hamiltonian by employing the Holstein-Primakoff transformation.
If we discuss the synchronization of such systems, once an underlying limit-cycle has been established, we could employ the measures reviewed earlier to propose a measure of synchronisation based in the relative phase.
On the other hand, if we wish to discuss the $N\approx1$ limit of the Tavis-Cummings model, then all previous measures fail.
In contrast to this, our new measure can handle the synchronization of such a dissimilar systems without any further modifications.

As an example of the synchronization of dissimilar quantum systems, consider a van der Pol oscillator coupled to an equally spaced 3-level system.
The Hamiltonian for the combined system in the rotated frame is given by
\begin{eqnarray}
    \hat{H}_R &=& \hat{H}_0 + \epsilon (\hat{S}_+ \hat{a} + \hat{a}^\dagger \hat{S}_-),
\end{eqnarray}
where the free Hamiltonian is $\hat{H}_0 = (\omega_a + \Delta) \hat{S}_z + \omega_a \hat{a}^\dagger \hat{a}$, $\omega_a$ is the natural frequency of the van der Pol oscillator, $\Delta$ is the detuning between the spin-1 system and the oscillator, and $g$ is the coupling strength.
The Lindblad master equation for the combined state $\rho$ is given by 
\begin{eqnarray}
    \dot{\rho} = -i[\hat{H},\rho] + \gamma_d \mathcal{D}(\hat{S}_-\hat{S}_z)\rho +\gamma_g \mathcal{D}(\hat{S}_+\hat{S}_z) \rho \nonumber\\
    +\Gamma_d \mathcal{D}(\hat{a}^2)\rho +\Gamma_g \mathcal{D}(\hat{a}^\dagger) \rho. 
    \label{eq:vdP_spin}
\end{eqnarray}
where $\Gamma_g(\gamma_g)$ and $\Gamma_d(\gamma_d)$ are the pumping and damping constants for the van der Pol and spin-1 atom, respectively.
Such a system represents a van der Pol oscillator coupled to a three-level system, both of which have been individually discussed in the literature before but never coupled.

The minimization of the relative entropy of synchronization in this case is performed over all possible diagonal limit-cycle states, and hence this example reduces to the discussion of Section {\ref{sec:diagonal}}.
We consequently obtain $\Omega_R = S_{\text{coh}}$, and this is plotted  as a function of detuning $\Delta$ and the coupling strength $\epsilon$ in Fig.\ref{fig:figure4}(b).

\subsection{Marginal State limit-cycles \label{sec:marginal}}

We now consider limit-cycle states which are not necessarily diagonal in the energy eigenbasis.
In order to obtain a closed-form expression for $\Omega_{\text{R}}(\rho)$, we consider an example where the two limit-cycle states are uncorrelated.
The full limit-cycle state is then given by
\begin{equation}
    \sigma=\sigma^A\otimes\sigma^B,\quad\text{where}\quad\sigma^{\alpha}=\sum_{ij}q_{ij}^{\alpha}|E_i^{\alpha}\rangle\langle E_j^{\alpha}|.
    \label{eq:marginal_state}
\end{equation}
Substituting this state into (\ref{eq:inequality}),
\begin{equation}
    \Omega_{\text{R}}(\rho) = \min_{\sigma} \left\{ -S(\rho) - \sum_{\alpha} \text{Tr} \left[ \rho \log \sigma^{\alpha} \right] \right\}
\end{equation}
Adding and subtracting the von Neumann entropies of the marginal states of $\rho$,
\begin{equation}
    \Omega_{\text{R}}(\rho) = \min_{\sigma^A,\sigma^B} \left\{ -S(\rho) + \sum_{\alpha} \left[ S(\rho^{\alpha}) + S\left( \rho^{\alpha} || \sigma^{\alpha} \right) \right] \right\}
\end{equation}
As $S(\rho^{\alpha}||\sigma^{\alpha})\geq 0$, the minimum value is obtained when $\sigma^{\alpha}=\rho^{\alpha}$, yielding,
\begin{equation}
    \Omega_{\text{R}}(\rho) = -S(\rho) +S(\rho^A) + S(\rho^B).
    \label{eq:mutual_inf}
\end{equation}
This shows that for the set of limit-cycle states given by uncorrelated non-diagonal states the closest states to the steady state $\rho$ are given by its marginals.
In turn, this means that the relative entropy of synchronization $\Omega_{\text{R}}(\rho)$ is given by the mutual information between the two systems.

This result is consistent with \cite{ameri2015mutual}, where the authors proposed mutual information as an order parameter for two mutually coupled van der Pol oscillators.
However, note that the mutual information was postulated in \cite{ameri2015mutual} to be a suitable measure by showing that is shared qualitative features with a physically motivated measure of complete synchronization.
We formalize this in Eq.~(\ref{eq:mutual_inf}) by showing that mutual information is the only possible measure when minimizing over set of limit-cycle states in Eq.~(\ref{eq:marginal_state}).

\subsection{Partially Coherent limit-cycle States \label{sec:partially}}

We now consider a new category of limit cycle states that have not been considered in literature before, namely limit-cycle states that are partially coherent.
This interpolates between diagonal limit-cycle states, where there are no off-diagonal terms present and marginal limit-cycle states, where all off-diagonal terms in the marginal density matrix could be non-zero.
This intermediate family of limit-cycle states allows us to discuss, for example, locally driven systems which are then coupled to each other.
Consider two spin-1 atoms which are locally driven across the $\ket{E_2}\bra{E_3}$ transition.
A generic representation of such a qutrit is written as $\sigma = \sum_i q_i |E_i \rangle \langle E_i|+ q_{23} |E_2\rangle \langle E_3|+ q_{32} |E_3\rangle \langle E_2|$ and constitutes our intermediate limit-cycle state.
If we now couple these two qutrits, our new measure can reveal any subsequent mutual synchronization of systems without including any possible entrainment due to these local drives.

Consider a diagonal marginal state, $\lambda =\sum_{i} \lambda_i |i\rangle \langle i|$. By applying a general unitary which acts on the subspace $\big\{|E_2\rangle,\; |E_3\rangle \big\}$, a general partially coherent limit-cycle state can be generated as $\sigma^{\alpha} = U \lambda U^\dagger = \sum_{k=1}^{3} q_k U|k \rangle \langle k|U^\dagger$, where $U = 1\oplus u$ and the unitary $u(\theta_1,\theta_2,\theta_3)$ which acts on the subspace $\big\{|E_2\rangle,\; |E_3\rangle \big\}$ is given by,
\begin{eqnarray} 
u(\theta_1,\theta_2,\theta_3) &=& e^{-i\theta_1 \hat{\sigma}_z}e^{-i\theta_2\hat{\sigma}_x}e^{-i\theta_3 \hat{\sigma}_z} \nonumber\\
&=& \left(\begin{array}{cc}
	e^{-i(\theta_1+\theta_3)}\cos \theta_2 & -i e^{-i(\theta_1-\theta_3)}\sin\theta_2  \nonumber\\
  -i e^{i(\theta_1-\theta_3)}\sin\theta_2 & e^{i(\theta_1+\theta_3)}\cos \theta_2
	\end{array}\right).
\end{eqnarray}

Again we can calculate, 
\begin{eqnarray}
    \omega_{\alpha} &=& \max_{p_k, u} \sum_k \log q_k^{\alpha}\;\langle k|\left(\rho^{\alpha}\right)'|k\rangle \nonumber\\
    &=& \max_{q_k, u} \big\{\rho_{11}^{\alpha}\log q_1^{\alpha}+ (\rho_{22}^{\alpha})' \log q_2^{\alpha} \nonumber\\
    & & + (\rho_{33}^{\alpha})' \log q_3^{\alpha}\big\},
\end{eqnarray}
where $\left(\rho^{\alpha}\right)'=u^\dagger \rho^{\alpha} u$. This gives,
\begin{eqnarray} 
    \omega_{\alpha} &=& \max_{q_k^{\alpha},\theta_1,\theta_2} \Big\{ \rho_{11}^{\alpha} \log q_1^{\alpha} \nonumber\\ 
    & & + \sin^2(\theta_2)\; (\rho_{22}^{\alpha} \log q_3^{\alpha} +\rho_{33}^{\alpha}\log q_2^{\alpha}) \nonumber\\
    & &+ \cos^2(\theta_2)\; (\rho_{22}^{\alpha}\log q_2^{\alpha} + \rho_{33}^{\alpha} \log q_3^{\alpha}) \nonumber\\
    & & -\sin 2\theta_2 \log \frac{q_2^{\alpha}}{q_3^{\alpha}} \text{Im} [e^{i2\theta_1}\rho_{23}^{\alpha}]\Big\}, \nonumber\\
    &=& \rho_{11}^{\alpha} \log \rho_{11}^{\alpha} \nonumber\\
    & & + \max_{q_2^{\alpha},\theta_2}\Big\{ \sin^2\theta_2\; (\rho_{22}^{\alpha} \log (1-\rho_{11}^{\alpha}-q_2^{\alpha}) \nonumber\\
    & & +\rho_{33}^{\alpha}\log q_2^{\alpha}) + \cos^2(\theta_2)\; (\rho_{22}^{\alpha}\log q_2^{\alpha} \nonumber\\
    & & + \rho_{33}^{\alpha} \log (1-\rho_{11}^{\alpha}-q_2^{\alpha})) \nonumber\\
    & & +\sin 2\theta_2 \; \log \frac{q_2^{\alpha}}{1-\rho_{11}^{\alpha}-q_2^{\alpha}} |\rho_{23}^{\alpha}|\Big\}.
\end{eqnarray}
where the optimal values for $q_1^{\alpha} = \rho_{11}^{\alpha}$ and $\theta_1^{\alpha} = \pi/2 - \phi_{23}^{\alpha}/2;\; \rho_{23}^{\alpha} = |\rho_{23}^{\alpha}| e^{i\phi_{23}^{\alpha}}$ are substituted in the above equation.
The final expression for the relative entropy of synchronization can be expressed as,

\begin{eqnarray}
    \Omega_{\text{R}} &=& -S(\rho) -\omega_A - \omega_B.\label{eq:partial}
\end{eqnarray}

We plot Eq.~(\ref{eq:partial}) in Fig.~\ref{fig:figure4}(c).
The Arnold tongue is now smaller when compared to the case of coupled spin-1 atoms without local drives in Fig.~\ref{fig:figure4}(a), indicating weaker mutual synchronization between the atoms.
This is expected as the atoms get entrained to their local drives.
In order for the atoms to become mutually synchronized they must now be more strongly coupled to overcome the effect of local entrainment.
The inset of Fig.~\ref{fig:figure4}(c) shows the difference between $\Omega_{\text{R}}(\rho)$ for partially-coherent limit cycle states and mutual in formation between the spin-1 atoms.

\section{Discussion \& Conclusions \label{sec:discussion}}

In this manuscript, we introduced a measure of synchronization based on distance to the limit-cycle dynamics.
This measure is inspired simultaneously by well known information theoretic measures of entanglement, discord and other quantum correlations and by an understanding of synchronization as the deviation from limit-cycle dynamics.
Furthermore, it is often desirable to study synchronization to a single drive in an otherwise complex system, and our measure allows such flexibility by the choice of the limit-cycle dynamics.
We capture synchronization and entrainment dynamics of a variety of different systems such as unipartite systems being entrained to external drives, bipartite or multipartite systems that are mutually coupled and driven quantum systems coupled to each other.
Our measure allows us to filter the dynamics we are interested in and construct a unified measure of synchronization that applies to finite and infinite dimensional systems identically.
We note that since the underlying set of limit-cycle states can be different for different examples, the synchronization measure of two completely disparate systems should not be directly compared.
Instead, by making sure that the underlying principles of the measure are the same, we expect that proportional changes in the measure of synchronization are meaningful to discuss.

This measure can be viewed as a generalization of the mutual information theoretic measure introduced in \cite{ameri2015mutual}  and hence we comment on critical differences.
Firstly, since mutual information is a bipartite measure, it does not accommodate unipartite systems that are entrained to external signals.
Likewise, since strong sub-additivity inequality does not hold for generic multipartite systems, the mutual information theoretic measures cannot be applied to multipartite systems.
In contrast to this, our relative entropy is always evaluated between the steady state and the ``limit-cycle" state, our measure is well defined.
The relative entropy can be infinite if the support of $\rho_{\text{lim}}$ is different than the support of $\rho$, but this is easily remedied by moving to the trace distance measure of synchronisation.
Furthermore, mutual information does not allow us to capture complex dynamical systems which involve coupling of locally driven quantum systems.
Finally, we note that if we define the limit-cycle states to be the tensor product of the marginal states, we recover the mutual information measure. 

Likewise, our measure reduces to the relative entropy of coherence if the underlying set of limit-cycle states is chosen to be the set of diagonal states.
This conceptually generalizes the relationship between synchronization and $l_1$ norm of coherence presented in \cite{jaseem2020quantum}.
Our measure indeed interpolates between the two scenarios, one where nothing is known about the subsystem dynamics (and hence the marginal states are chosen as the limit-cycle states) and another where a lot is known about the limit-cycle dynamics (that they are diagonal).
By interpolating between these two situations, we demonstrate synchronization between dissimilar systems not considered before.
In Section \ref{sec:dissimilar}, we investigated mutual synchronization between a three-level atomic degree of freedom and a van der Pol oscillator.
Likewise, in Section \ref{sec:partially}, we considered another novel example and studied mutual synchronization between two partially coherent three-level atoms.
We hope that our measure will lead to applications of quantum synchronization in hybrid quantum systems such as opto- and nano-mechanical systems, vacancy centers coupled to cavities and circuit QED. 


\begin{acknowledgments}
SV acknowledges support from an IITB-IRCC grant number 16IRCCSG019, and a DST-SERB Early Career Research Award (ECR/2018/000957). MH acknowledges support by the Air Force Office of Scientific Research under award number FA2386-19-1-4038.
\end{acknowledgments}
 

%

\end{document}